\documentclass[prb,twocolumn,aps,showpacs]{revtex4}
\usepackage{graphicx,bm,hyperref,amsmath,amssymb}
\newcommand{\mscr}[1]{{\mbox{\scriptsize #1}}}
\begin{document}
\title{
Evolution of a localized electron spin in a nuclear
spin environment}
\author{Sigurdur I.\ Erlingsson}
\altaffiliation[Present address: ]{Department of Physics and Astronomy,
  University of Basel, Klingelbergstrasse 82, CH-4056, Switzerland}
\author{Yuli V.\ Nazarov} 
\affiliation{Delft University of Technology, Department of NanoScience,
         Lorentzweg 1, 2628 CJ Delft, The Netherlands}
\date{\today}
\begin{abstract}
Motivated by recent interest 
in the role of the hyperfine interaction in
quantum dots we study the dynamics 
of a localized electron spin
coupled to many nuclei. 
An important feature of the model
is that the coupling to an individual nuclear spin
depends on its
position in the quantum dot. 
We introduce a semi-classical description of the system 
valid in the limit
of a large number of nuclei and analyze 
the resulting classical dynamics.
Contrary to a natural assumption,
the correlation functions of electron spin with
an arbitrary initial 
condition show no decay in time. 
Rather, they exhibit complicated undamped oscillations.  
This may be attributed to the fact that the system has many
integrals of motion and is close to an integrable one.
The ensemble averaged correlation
functions do exhibit a 
slow decay $(\sim 1/\ln(t))$ for $t \rightarrow \infty$.
\end{abstract}
\maketitle
\section{Introduction}
The coherent manipulation of localized spins in solid state systems is
currently a very active field research\cite{awschalom02:xx}.  
One of the most
ambitious goals in this field of research is developing a quantum bit,
or qubit.  Such qubits would form the basic building blocks of quantum
computers\cite{steane98:117,shnirman97:2371,loss98:120,kane98:133}. 
The strength of the qubit coupling to 
the environment should  be sufficiently well controlled so that the
residual environmental interactions are below a threshold set by the
operating speed of the
device\cite{nakamura99:786,mooij99:1036,vandersypen01:883}.  

Since the first proposal in Ref.\ \onlinecite{loss98:120}, the qubit
candidate based on localized electron spins in quantum dots has been
the subject of many theoretical and experimental studies.  
There have already  
been quite a few studies of the relaxation time of
electron spins in GaAs 
quantum dots.
The role of various spin-orbit related mechanisms has been considered
for spin-flip transitions between singlet and
triplets\cite{khaetskii00:12639} and Zeeman split
doublets\cite{khaetskii01:125316}. The hyperfine mediated spin-flip
rates were investigated for singlet-triplet\cite{erlingsson01:195306}
and doublet\cite{erlingsson02:155327} transitions.  A novel spin-flip
mechanism due to interface motion has also been
investigated\cite{woods02:161318R}  and
Coulomb blockade effects in the nuclear spin relaxation time were
looked at in Ref.\ \onlinecite{lyandaGeller02:107602}.
For GaAs quantum dots measurements of spin relaxation times, or $T_1$, between
singlet-triplet states\cite{fujisawa01:081304R,fujisawa02:278} and
doublet\cite{hanson03:196802} states have been done.  The measurements
give $T_1 \approx 200$\,$\mu$s (singlet-triplet) and $T_1 >
50$\,$\mu$s (doublet).  For the doublet measurements only a lower bound
for the relaxation time is obtained \cite{hanson03:196802}, so
comparison to various theoretical calculations remains difficult. 

The operation of a qubit requires a coherent superposition of
states that persists during the time of the qubit operation.
Several papers have
addressed the question of the decoherence time of electron spin in
GaAs quantum
dots.
The decoherence of a single electron spin may be caused by inhomogeneous
hyperfine interaction\cite{khaetskii02:186802,schliemann02:245303} but
when an ensemble of electron spins is considered the timescale of
decoherence is determined by the fluctuations in the total field
which acts on the electron spin due to the
nuclei\cite{merkulov02:205309,semenov03:73301}.  
The role of fluctuations in the nuclear spin system, socalled spectral
diffusion, on the electron spin decoherence has also been
studied\cite{deSousa03:33301,deSousa03:115322}.

In Ref.\ \onlinecite{khaetskii02:186802} the characteristic timescale for
the decay of a 
specific correlation function was associated with the decoherence 
time of the electron spin.  The decohence in this case was attributed
to the 
spatially dependent hyperfine coupling constant which caused small
frequency changes through flip-flop processes involving spatially
separated nuclei.  
A different approach was used in Ref.\ \onlinecite{merkulov02:205309}
which relied on representing the nuclear spin system as 
an effective nuclear magnetic field that couples to the electron spin
in the quantum dot.  
Merkulov {\em et al}.\ discussed some basic features of this
semi-classical approach but the
calculations were done for ensemble averaged
quantities\cite{merkulov02:205309}. 

In the present work we combine the approaches mentioned
and extend the semiclassical 
effective-field method
\cite{erlingsson02:155327}
to include the effects of the spatially varying hyperfine coupling
constant.
Due to the large difference of 
timescales for the electron and nuclear spin systems we are able to
solve the problem in two steps.
In the first step we establish that the
nuclear system can be treated as an adiabatic effective nuclear
magnetic field acting on the electron.
The latter step involves 
the back action of the electron spin which will determine the  
evolution of the nuclear spins.
The fact that the single electron spin is coupled to a large number of
nuclei in the quantum dot, $N$, but each nucleus is only
coupled to the single electron spin hints 
at an asymmetry in the behavior of the electron and nuclear spins.
The electron precesses in  an effective nuclear magnetic field
which is due to the whole nuclear spin system.  
This field is $\propto \sqrt{N}$ times larger than the
corresponding hyperfine field due to the single electron
spin in which the nuclear spins precess.
Thus the dynamics of the
electron are much faster than the dynamics of the nuclei.
Also, the large number of nuclei involved makes it possible to treat
the nuclear system in a semi-classical
way\cite{dyakonov86:110,merkulov02:205309,erlingsson02:155327}.  We
do not include the dipole-dipole interaction of the nuclei, 
since the effects of such interaction can only become noticeable
at very long timescale 
( $t_\mscr{dip}\simeq 10^{-3}$sec, for GaAs quantum dots
\cite{merkulov02:205309,khaetskii03:195329}).  

The resulting dynamics are represented by a set of
equations of motion for several subsystems  of
nuclear spins. 
Each subsystem is characterized by the same value of coupling to
the electron spin. 
The dynamical equations are non-linear and comprise many degrees 
of freedom.  From this, one generally expects
chaotic and ergodic dynamics,
so that the memory about initial
conditions is lost at a certain time scale. 
This would result in the decay of correlation functions
at this time scale and was an initial motivation of this research.
However, we prove that the actual dynamics is not chaotic.
The calculated correlation functions show complex, yet
regular, oscillations that persist for long timescale without
discernible decay. We explain this by conjecturing that the system has
many integrals of motion so it is close to
exactly integrable one.  

If we average the dynamics over all possible 
initial conditions, the averaged correlation function 
does show decay in
time. For large times, it is  inversely proportional to logarithm of time.

The remaining text of the paper is organized as follows.  In Sec.\
\ref{sec:hyperfine} the hyperfine interaction in a quantum dot is
presented and its semi-classical representation discussed.
The adiabatic approximation for the electron spin dynamics is discussed
in Sec.\ \ref{sec:adiabatic} and some analytic properties of the
dynamical equations are presented in Sec.\ \ref{sec:analytic}.  
The correlation functions
used to characterize the dynamics are introduced in Sec.\
\ref{sec:correlationFunction} and finally the results of the numerical
calculations are discussed in Sec.\ \ref{sec:results}.
\section{Hyperfine interaction in quantum dots and its semi-classical
representation} 
\label{sec:hyperfine}
The Hamiltonian describing the hyperfine coupling between conduction
band electrons and the lattice nuclei in GaAs is of the well known
form of the contact potential 
\begin{equation}
H_\mscr{HF}=A\hat{\bm{S}}\cdot \sum_k \hat{\bm{I}}_k
\delta(\bm{r}-\bm{R}_k),
\label{eq:hyperfineHamiltonian}
\end{equation}
where $A$ is the hyperfine constant, $\hat{\bm{S}}$ ($\hat{\bm{I}}_k$)
and $\bm{r}$ ($\bm{R}_k$) are respectively the spin and position of
the electron ($k$th nuclei).
In the GaAs conduction band (which is mainly composed of
$s$-orbitals), the dipole-dipole part of the hyperfine 
interaction vanishes \cite{brown98:49}.
In this paper the focus will be on electrons localized in a quantum
dot, and the hyperfine interaction in such systems.
The quantum dots considered here are quite general, but we introduce
some restrictions to simplify the model.
First of all it is assumed that the number of electrons is fixed,
preferably to one.  From the experimental point of view
this assumptions is quite reasonable since 
having only a single electron in the dot has
already been demonstrated
experimentally\cite{tarucha96:3613,elzerman03:161308}.   
The second assumption is that the orbital level splitting is much
larger than the hyperfine energy.  
In this case the hyperfine Hamiltonian can be projected to the lowest
orbital level since contributions from higher orbitals are strongly
suppressed due to the large orbital energy separation.
If the ground state orbital $\psi(\bm{r}$) of the
quantum dot is known, then an effective spin Hamiltonian can be
written as 
\begin{equation}
H_\mscr{s}=g\mu_B \bm{B}\cdot\hat{\bm{S}}
+ \gamma_\mscr{GaAs} \sum_k\bm{B}\cdot \hat{\bm{I}}_k+
\sum_k A |\psi(\bm{R}_k)|^2 \hat{\bm{S}}\cdot \hat{\bm{I}}_k
\label{eq:hamiltonian}
\end{equation}
where $g$ is the $g$-factor, $\mu_B$ is the Bohr magneton,
$\gamma_\mscr{GaAs}$ is the gyromagnetic ratio of the effective nuclear
species and $\bm{B}$ is the external applied field.

In a typical quantum dot a
single electron spin may be coupled to $10^4-10^6$ nuclear spins.  
When the electrons interacts with so many nuclei it is
possible to interpret the combined effect of the nuclei as an
effective magnetic field.  
Before proceeding further it is convenient to introduce a different
way of writing the hyperfine interaction in the last term in Eq.\
(\ref{eq:hamiltonian}). 
The wave function of the ground state orbital has some characteristic
spatial extent which is determined by the confining potential.  Without loss
of generality it may be assumed that the lateral (with respect to the
underlying 2DEG) and transverse confining lengths are $\ell$ and $z_0$
respectively.   
Defining the volume of the quantum dot as $V_\mscr{QD}=\pi z_0
\ell^2$, a dimensionless function is introduced:
\begin{equation}
f(\bm{R}_k)=V_\mscr{QD} |\psi(\bm{R}_k)|^2.
\label{eq:f}
\end{equation}
Furthermore, denoting the maximum value of $f$ with $f_\mscr{Max}$ we
introduce a dimensionless coupling constant
\begin{equation}
g_k=g(\bm{R}_k)=\frac{f(\bm{R}_k)}{f_\mscr{Max}} \in (0,1).
\label{eq:gk}
\end{equation}
The hyperfine coupling constant may be expressed in terms of the
concentration, $C_n$, of nuclei with spin $I$ and a characteristic
energy $E_n$ through the relation $A=E_n/C_n I$.
The energy $E_n$ is the maximum Zeeman splitting possible due to a fully
polarized nuclear system, its value being $E_n\approx0.135$\,meV in
GaAs\cite{paget77:5780,dobers88:1650}. The hyperfine interaction term
can thus be written as
\begin{equation}
H_\mscr{HF}=\hat{\bm{S}}\cdot\hat{\bm{K}},
\end{equation}
where we have introduced the operator for the effective nuclear
magnetic field
\begin{eqnarray}
\hat{\bm{K}}
&=&\gamma \sum_k g_k \hat{\bm{I}}_k,
\label{eq:Kop2}
\end{eqnarray}
and the characteristic hyperfine induced nuclear spin precession frequency
\begin{equation}
\gamma=\frac{E_n f_\mscr{Max}}{NI}.
\end{equation}
As was shown in Ref.\ \onlinecite{erlingsson02:155327} it is possible
to replace the operator in Eq.\ (\ref{eq:Kop2}) with a effective
nuclear magnetic field $\bm{K}$.
Its initial value is random since it is determined
by unknown details of the nuclear system.  
Assuming that the nuclear spin system temperature $k_BT \gg \gamma$,
all nuclear states are equally likely to be occupied.  Initially the
nuclei are nominally decoupled from each other and the distribution of
$\bm{K}$ is to a good approximation represented by a  
Gaussian\cite{erlingsson02:155327,merkulov02:205309} 
\begin{equation}
P(\bm{K})=\left (\frac{3}{2 \pi \gamma^2
N\overline{I^2} } 
\right )^{3/2}  
\exp
\left (
 -\frac{ 3\bm{K}^2}
                   {2 \gamma^2N\overline{I^2}} 
\right ),
\label{eq:Kweight}
\end{equation}
where $\overline{I^2}=I(I+1)N^{-1}\sum_kg_k^2$.

In the above discussion the dynamics of the nuclear spin system was
disregarded.  
Before we include its dynamics it is instructive to
first derive {\em exact} operator equations of motion, and then apply
the semi classical approximation to those equations.
The dynamics of the combined electron and nuclear spin systems are
determined by the Heisenberg equation of motion 
\begin{eqnarray}
\frac{d}{dt}\hat{\bm{S}}&=&\hat{\bm{K}}\times \hat{\bm{S}} +
g \mu_B \bm{B} \times \hat{\bm{S}},
\label{eq:dS}\\
\frac{d}{dt}\hat{\bm{I}}_k&=&
\gamma
g_k \hat{\bm{S}}\times
\hat{\bm{I}}_k +\gamma_\mscr{GaAs} \bm{B} \times \hat{\bm{I}}_k.
\label{eq:dIk}
\end{eqnarray}
Multiplying Eq.\ (\ref{eq:dIk}) with $g_k$ and summing over $k$ gives 
the equation of motion for $\hat{\bm{K}}$ 
\begin{eqnarray}
\frac{d}{dt}\hat{\bm{K}}=
\gamma
\hat{\bm{S}} \times \left ( \sum_k g_k^2 \hat{\bm{I}}_k \right )
+\gamma_\mscr{GaAs} \bm{B} \times \hat{\bm{K}}.
\label{eq:dK}
\end{eqnarray}
In contrast to the
simple dynamics of Eq.\ \ref{eq:dS}, the
equation of motion for $\hat{\bm{K}}$ is quite complicated.  
The reason for the asymmetry 
is the position dependent coupling $g_k$.
The quantity in the brackets on the rhs of Eq.\ (\ref{eq:dK}) cannot
be expressed in terms of $\hat{\bm{K}}$.
Only in the simple case of constant $g_k$ it is possible to
write a closed equation of motion for $\hat{\bm{K}}$ and
$\hat{\bm{S}}$.\cite{semenov03:73301}
Without actually solving Eqs.\ (\ref{eq:dS}) and (\ref{eq:dK}) it is
still possible  
to extract  general features of the dynamics.
For zero external magnetic field, the electron spin will precess with
frequency $\propto E_n N^{-1/2}$ (which is the magnitude of the
effective nuclear magnetic field) and the nuclear system precesses with
frequency $\propto E_n N^{-1}$.  
Thus, for $N \gg 1$,
the electron spin effectively sees a 
stationary 
nuclear system and in turn the nuclear system sees a time averaged
electron spin. 

To incorporate $(i)$ the separation of timescales and $(ii)$ the
inhomogeneous coupling we introduce a scheme that separates the nuclear system
into $N_b$ subsystems, each being characterized by a fixed 
coupling $g_b$.
The effective nuclear magnetic field of a given subsystem is
\begin{equation}
\hat{\bm{K}}_b=
\gamma
g_b\sum_{k\in b} \hat{\bm{I}}_k
\end{equation}
where the notation $k\in b$ is shorthand for all nuclei whose
coupling is $g_k \in [g_b-\delta g/2,g_b+\delta g/2]$, with $\delta
g=1/N_b$ being the coupling constant increment.
As long as $N_b \ll N$ each subsystem can be replaced by a
classical variable $\hat{\bm{K}}_b\rightarrow \bm{K}_b$, which
represents the effective nuclear field due to that particular nuclear
spin subsystem.  

Using the same procedure as was used in deriving Eq.\ (\ref{eq:dK}) we
arrive at a equation of motion for $\hat{\bm{K}}_b$ and applying the
semi-classical approximation results in
\begin{eqnarray}
\frac{d\bm{K}_b}{dt}= 
\gamma
g_b \langle \bm{S}\rangle \times \bm{K}_b 
+\gamma_\mscr{GaAs} \bm{B} \times \bm{K}_b,
\label{eq:dKb}
\end{eqnarray}
where $\langle \bm{S} \rangle $ is
an appropriate time averaged electron spin.  As will be discussed in
the next section, this average electron spin may be written as a
function of the total effective nuclear field
\begin{equation}
\bm{K}=\sum_b \bm{K}_b.
\end{equation}
The initial condition for each nuclear spin subsystem is randomly
chosen from a Gaussian distribution whose variance is 
(see Appendix \ref{app:variance}) 
\begin{eqnarray}
\langle \bm{K}_b^2 \rangle &=&
\gamma^2 N I(I+1)g_b\delta g.
\label{eq:Kb2}
\end{eqnarray}
The set of differential equations in Eq.\ (\ref{eq:dKb}), with the
associated  random initial 
conditions constitute a set of autonomous differential equations.

Separating the nuclear system into subsystem  with a
constant $g_b$ is an approximation to the continuous coupling $g_k$.
As the number of  subsystems increases $g_b$ will more closely
represent $g_k$.  However, for the semi-classical approximation to
be valid each subsystem must contain many nuclear spins.  Thus,
increasing $N_b$ should better reproduce the actual system, as long as
$N_b\ll N$. 
\section{Adiabatic approximation for the electron spin}
\label{sec:adiabatic}
As we have shown in the previous section,
the nuclear spin system may be treated as a slowly varying
effective nuclear magnetic field acting on the electron spin.
Letting $\bm{H}(t)$
represents any slowly varying magnetic field (fulfilling the usual
adiabatic conditions) acting on a single 
electron spin, leads to the Hamiltonian 
\begin{equation}
H_e(t)= \hat{\bm{S}}\cdot \bm{H}(t).
\label{eq:H_e}
\end{equation}
It is convenient to introduce the instantaneous eigenfunctions of the
Hamiltonian, which are solutions of
\begin{equation}
H_e(t) |\bm{n}(t);\pm\rangle =E_\pm(t)  |\bm{n}(t);\pm\rangle .
\end{equation}
The eigenstates are labeled by $\bm{n}(t)$ to indicate that
these eigenstates are either pointing `up' (+) or `down' (-) along the
total magnetic field, whose direction is determined by the unit
vector 
\begin{equation}
\bm{n}(t)=\frac{\bm{H}(t)}{|\bm{H}(t)|}.
\end{equation}
For a spin $1/2$ in an external field, the eigenenergies are
$E_\pm(t)=\pm \frac{1}{2}|\bm{H}(t)|$, and the corresponding eigenstates are
written in the basis of 
the $\hat{S}_z$ eigenvectors 
\begin{eqnarray}
\left (
\begin{array}{l}
|\bm{n}(t);+\rangle \\
|\bm{n}(t);-\rangle
\end{array}
\right )
=
\frac{1}{\sqrt{1+|a(t)|^2}}
\left (
\begin{array}{l}
|\uparrow\rangle +a(t)|\downarrow\rangle \\
|\downarrow\rangle-a^*(t)|\uparrow\rangle
\end{array}
\right ).
\end{eqnarray}
The time dependent mixing of spin components 
is given by
\begin{equation}
a(t)=\frac{|\bm{H}(t)|-H_z(t)}{H_x(t)-iH_y(t)}.
\label{eq:a}
\end{equation}
The wave function may be expanded in basis of instantaneous eigenstates 
\begin{equation}
|\psi(t)\rangle=\sum_{\sigma=\pm}c_\sigma(t) |\bm{n}(t);\sigma\rangle 
\label{eq:psi}
\end{equation} 
where the expansion coefficients are
\begin{equation}
c_\pm(t)=c_\pm(t_0)e^{\left( 
i\gamma_\pm(t)-\frac{i}{\hbar}\int^t_{t_0} d\tau E_\pm(\tau)
\right)}.
\end{equation}
The additional phase factor appearing in the previous equation is the
usual adiabatic phase \cite{schiff68:1}
\begin{equation}
\gamma_\pm(t)=i\int^t_{t_0} d\tau  
\langle \bm{n}(\tau);\pm|d/d\tau |\bm{n}(\tau);\pm \rangle.
\end{equation}
Integrating by parts the rhs of the last equation and using the
orthogonality of the instantaneous eigenstates, it can
be shown that the 
phase $\gamma_\pm$ is a real number.  Also, for a doublet
$\gamma\equiv \gamma_+=- \gamma_-$.  

\begin{figure}[ht]
\begin{center}
\includegraphics[angle=0,width=5cm]{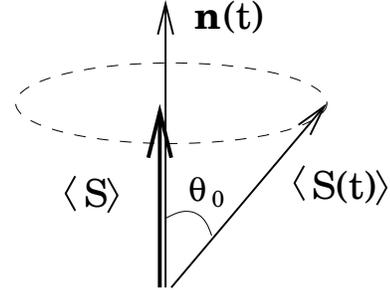}
\caption{The time dependent electron spin $\langle \bm{S}(t) \rangle$
precesses rapidly around the total effective 
magnetic field, resulting in a slowly varying average spin $\langle
\bm{S}\rangle$ that the nuclei see. The angle between the instantaneous
electron spin and $\bm{n}$ is $\theta_0$}
\label{fig:aveSpin}
\end{center}
\end{figure}

Using the wave function in Eq.\ (\ref{eq:psi}), the average electron
spin is 
\begin{eqnarray}
\langle \hat{\bm{S}}(t)\rangle&\equiv& \langle
\psi(t)|\hat{\bm{S}}|\psi(t)\rangle \nonumber \\
&=&\sum_{\sigma=\pm}|c_\sigma(t_0)|^2
\langle \bm{n}(t);\sigma|\hat{\bm{S}}|\bm{n}(t);\sigma\rangle
\nonumber \\
&+&2\Re \! \left \{ \! c_+^*(t_0)c_-(t_0)e^{\left( 
-2i\gamma(t)-\frac{i}{\hbar}\int^t_{t_0} d\tau |\bm{H}(\tau)| 
\right )} \! \right \}\!\!. 
\label{eq:aveSpin}
\end{eqnarray}
The latter term in Eq.\ (\ref{eq:aveSpin}) oscillates with frequency
$|\bm{H}(t)|/h \gg E_n N^{-1}/h$, so its average is zero on the
timescales of the nuclear system.  The average value of the electron
spin entering Eq.\ (\ref{eq:dKb}) is
\begin{eqnarray}
\langle \bm{S} \rangle &=& \sum_{\sigma=\pm}|c_\sigma(t_0)|^2
\langle \bm{n}(t);\sigma|\hat{\bm{S}}|\bm{n}(t);\sigma\rangle
\nonumber \\
&=&\frac{1}{2}\cos (\theta_0)
\frac{\bm{H}(t)}{|\bm{H}(t)|},
\end{eqnarray}
where $\theta_0$ is the angle between the initial electron spin and
$\bm{n}(0)$, see Fig.\ \ref{fig:aveSpin}.  
The orientation between $\bm{S}$ and $\bm{n}(0)$ changes the
precession of all $\bm{K}_b$'s by a multiplicative factor
$\cos\theta_0$.   
This overall factor has no effect on the dynamics and we subsequently
put it to unity. Physically, one
might think of a dissipation mechanism that would initially align the
electron  spin and $\bm{H}$ to each other, although the mechanism
itself is not critical for the following discussion.

From the semi-classical version of Eq.\ (\ref{eq:dS}), the slowly
varying magnetic field is $\bm{H}(t)=\bm{B}+\bm{K}(t)$ which results in an
equation for $\langle \bm{S}\rangle$ that depends only on $\bm{K}$ and
$\bm{B}$.
Also, since we assume that the quantum dot is initially in the
ground state orbital and that the orbital energy separation is much
larger than the hyperfine energy,
there are no `Rabi oscillations' to higher orbitals.
\section{Some analytic properties}
\label{sec:analytic}
The rest of the paper will only consider the case of small magnetic
field $g\mu_B\bm{B}\ll  \bm{K}$, which is the regime where the
dynamics are most interesting.
In the opposite situation the magnetic field strongly constraints all
dynamics.   
The average electron spin, for $\bm{B}=0$ is
\begin{equation}
\langle \bm{S}\rangle=-\frac{1}{2}\frac{\bm{K}}{|\bm{K}|}
\end{equation}
and the resulting equation of motions for the nuclear spin
subsystems are
\begin{equation}
\frac{d}{dt}\bm{K}_b= \tilde{\gamma} g_b
\bm{K}\times \bm{K}_b,
\label{eq:dKb_calc}
\end{equation}
where $\tilde{\gamma}=-\gamma/2|\bm{K}|$, since
$|\bm{K}|$ is a constant of motion
\begin{equation}
\frac{d}{dt}|\bm{K}|^2=0, 
\end{equation}
and it equally changes the precession frequency of all the block
$\bm{K}_b$. 
In addition to $|\bm{K}|$, more integrals of
motion can be constructed from Eq.\ (\ref{eq:dKb_calc}):
\begin{eqnarray}
0&=&\frac{d}{dt}|\bm{K}_b|^2 \\
0&=&\frac{d}{dt}\bm{I}=\frac{d}{dt}
\left (\sum_b \frac{\bm{K}_b}{g_b}
\right )
\label{eq:I}\\
0&=&\frac{d}{dt}\left ( \bm{I}\cdot 
\left (
\sum_b \frac{\bm{K}_b}{g_b^2}
\right )
\right ).
\end{eqnarray}
The integral of motion in Eq.\ (\ref{eq:I}), is actually the total
spin $\bm{I}$ of the nuclear system\cite{merkulov02:205309}.
The integrals of motion are expected to affect the dynamics,
i.e.\ the system will be  `constrained' by them.

The solution for the electron dynamics is determined by the
dynamics of the nuclear system, which is encapsulated in $\bm{K}(t)$.
Although  
the dynamics are complicated there are some
ways to characterize the motion of the nuclear system.  For example, we can
define the following quantity 
\begin{equation}
\bm{K}(t;\zeta)=\sum_b\frac{1}{1+\gamma_b\zeta}\bm{K}_b(t)
\end{equation}
which satisfies the following equation of motion
\begin{equation}
\frac{d}{dt}\bm{K}(t;\zeta)=\frac{\bm{K}(t;0)\times
  \bm{K}(t;\zeta)}{\zeta}. 
\end{equation}
It is possible to construct other similar equations, but in general no
simple solution for them exist.
\section{Correlation functions}
\label{sec:correlationFunction}
A wide class of classical systems 
exhibits decaying correlation functions.  This occurs in classically 
chaotic systems in which the motion is such that the memory about initial
conditions is lost at some typical timescale \cite{ozorio88:xx}.
Most of sufficiently complicated classical systems are eventually
chaotic.  One might expect that the set of equations Eq.\
(\ref{eq:dKb}) should describe chaotic dynamics and decay of
correlation functions.  We will see later on that this is not the case.  

A useful way to characterize the electron spin dynamics it
is to introduce certain correlation functions.  For an isolated
quantum system these correlation function are expected to oscillate
periodically, without any decay.  
Incorporating environmental effects usually shows up in modified
behavior of the correlation functions.  The expected behavior is
that they should decay as a function of time.  
To investigate how the nuclear spin system acts as
a spin bath (environment), we introduce the following correlation
function
\begin{eqnarray}
G(t)&=&
\langle \uparrow
|\hat{\bm{S}}(t)\cdot\hat{\bm{S}}|
\uparrow\rangle,
\end{eqnarray}
where the time evolution of the operators is in the usual Heisenberg
picture. 

Since we are focusing on the slow dynamics it is useful to
write these correlation functions for long timescales.
In the adiabatic approximation the correlation functions may be
written as
\begin{eqnarray}
G(t)&=&
\frac{1}{4} 
\frac{(1-|a(t)|^2)(1-|a(0)|^2) +4a(t)a^*(0)}
{(1+|a(t)|^2)(1+|a(0)|^2)},
\label{eq:G0}
\end{eqnarray}
where the $a$'s are defined in Eq.\ (\ref{eq:a}).
The most interesting regime corresponds to weak external magnetic
fields. In this case there is no preferred direction and the dynamics 
show the richest behavior. 
In that limit the correlation functions take the simplified form 
\begin{equation}
G(t)=\frac{\bm{K}(t)\cdot \bm{K}(0)}{4K(t) K(0)}
\label{eq:G_B0}
\end{equation}
From these equations it is evident that the electron spin
correlation function is determined by the nuclear system variables for
times $t \gg \hbar \sqrt{N}/E_n$.  

When dealing with many identical systems in which the electron can be
prepared in a given initial state but the effective nuclear magnetic
fields differ in the initial values, ensemble averaged correlation
function must be considered.
No information is available about
the state of the effective nuclear magnetic field, except that their
initial vales are Gaussian distributed.  In this case the correlation
function in Eq.\ (\ref{eq:G0})
should be averaged over the appropriate distributions
\begin{equation}
\langle G(t) \rangle
=\int \prod_b d\bm{K}_{b,0}
P(\{\bm{K}_{b,0}\}) 
G(t;\{\bm{K}_{b,0}\}),
\label{eq:Gave_B0}
\end{equation}
where $\bm{K}_{b,0}=\bm{K}_b(0)$ and $P$ is the Gaussian
distribution of the initial values.  
Note that the correlation
functions appearing in Eq.\ (\ref{eq:G_B0}) are also
functions of the set of initial conditions $\{\bm{K}_b(0)\}$.   

\section{Results}
\label{sec:results}
The correlation functions in Eqs.\ (\ref{eq:G_B0}) and
(\ref{eq:Gave_B0}) are in general not exactly solvable so numerical
simulations have to be used.  In order to calculate  them
time series for $\bm{K}_b(t)$ need to be calculated.
These are obtained by numerically integrating 
the differential equations in Eq.\ (\ref{eq:dKb_calc}).
We focus on the case of no external magnetic field.
In GaAs $\gamma \approx 10^{-7}$\,meV$\approx10^5$\,Hz for quantum
dots containing $N\approx10^6$ nuclei. 
The differential equations are solved by
integrating numerically Eq.\ (\ref{eq:dKb_calc}) using the 4th order Runge
Kutta method.  
The $K_b(t)$'s are then used to calculate  $a(t)$
that enter Eq.\ (\ref{eq:G0}).  

\begin{figure}[ht]
\begin{center}
\includegraphics[angle=-90,width=8cm]{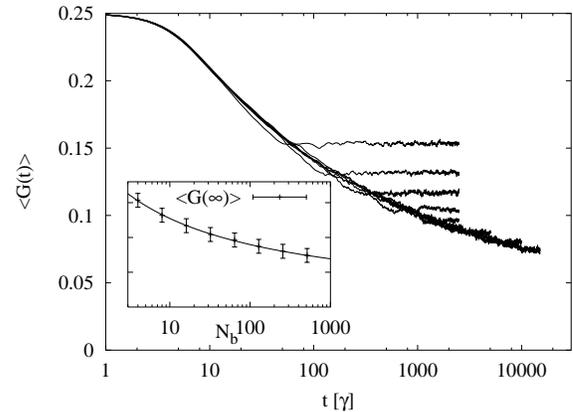}
\caption{The ensemble averaged correlation function as a function of
time for $N_b=4,8,16,32,64,128,256$ and 512.  The inset show the asymptotic
values of $\langle G(t)\rangle$ and a fit to $\alpha/\ln(\beta N_b)$.}
\label{fig:aveCorr}
\end{center}
\end{figure}

For the ensemble averaged correlation functions many sets of
time series $\{\bm{K}_b(t)\}$ are calculated, each corresponding to
different random initial condition chosen from a Gaussian distribution.  
The results for the ensemble averaged correlation function in Eq.\
(\ref{eq:Gave_B0}) are presented in Fig.\ \ref{fig:aveCorr}.  Each
curve is the result of calculations for different number of subsystems
$N_b=4,8,\dots,512$.  As is to be expected, the
correlation functions decay in time but a saturation value is reached
for sufficiently long times, which is determined by
$t_\mscr{sat} \propto \gamma^{-1}N_b$. 
This saturation is an artifact of the discretization, i.e.\ it
introduces a time above which the calculated correlation function no
longer represents the true correlation function. 
For the calculated correlation function to have a meaningful limit,
the saturation value should approach zero as $N_b$ increases and the
saturation time should go to infinity.  
The inset in Fig.\ \ref{fig:aveCorr} plots the saturation values of
$\langle G(t) \rangle $.  It is evident that the saturation values
tend to zero for larger $N_b$.  The decay fits an
inverse logarithm $\alpha/\ln(\beta N_b)$ quite well.  This
indicates that there is a well defined $N_b\rightarrow \infty$
correlation function which has an inverse logarithmic decay $\propto
1/\ln t$ as $t\rightarrow \infty$.  The origin of the logarithmic
decay is not 
fully understood (and thus the theoretical values of $\alpha$ and
$\beta$) but it might be related to the total spin 
$|\bm{I}|^2\propto \ln t_\mscr{sat}$ (see Appendix
\ref{app:variance}).  The correlation function is normalized and its
value might be dominated by $1/|\bm{I}|^2$ for $t>t_\mscr{sat}$.

\begin{figure}[ht]
\begin{center}
\includegraphics[angle=-90,width=8cm]{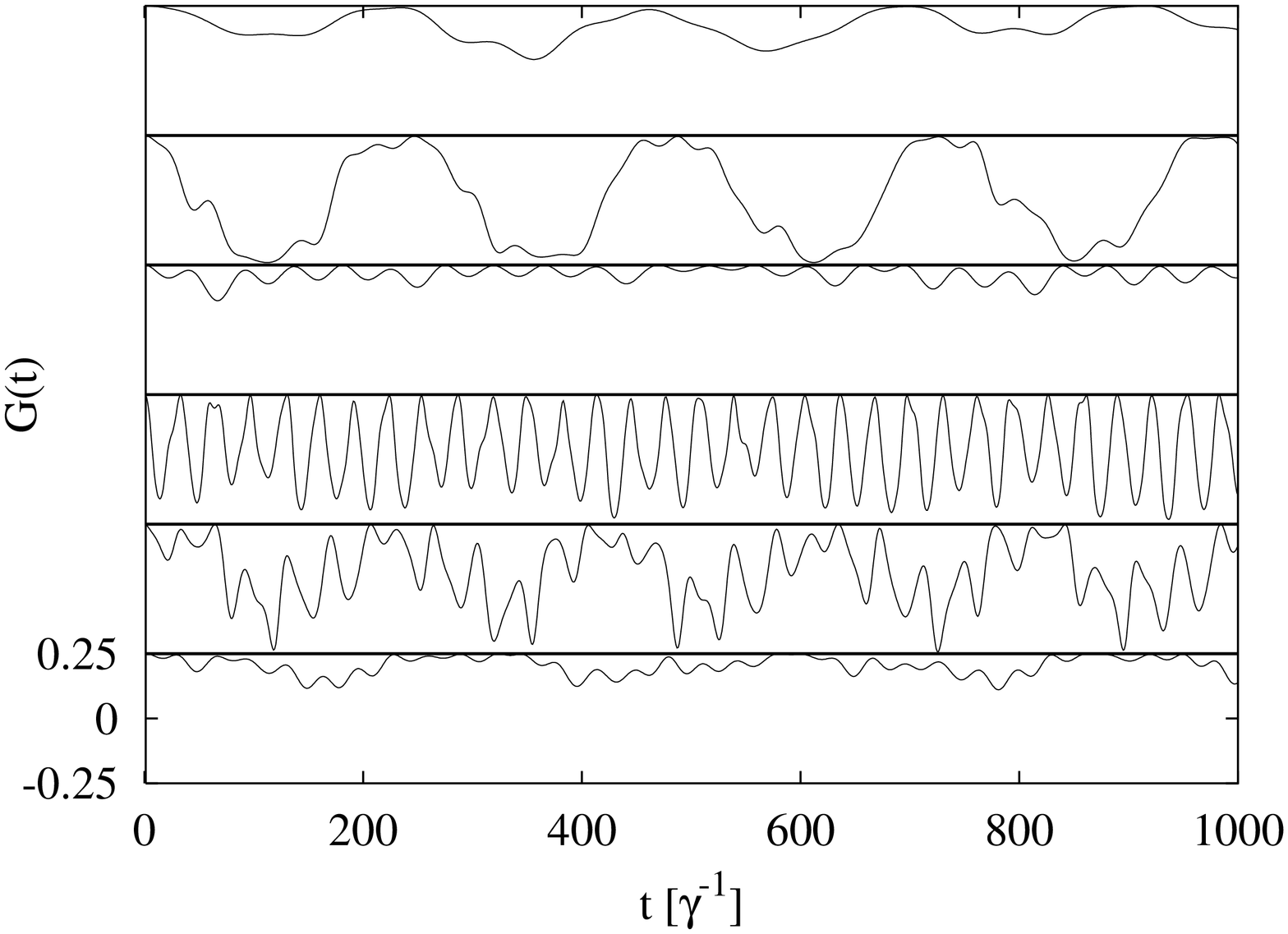}
\caption{Numerical calculations of the correlation function $G(t)$
for $N_b=8$ and various randomly chose initial conditions.  The curves are
offset for clarity and the vertical range is the same for all curves,
i.e.\ $-0.25$ 
to $0.25$} 
\label{fig:Nb8}
\end{center}
\end{figure}

\begin{figure}[ht]
\begin{center}
\includegraphics[angle=-90,width=8cm]{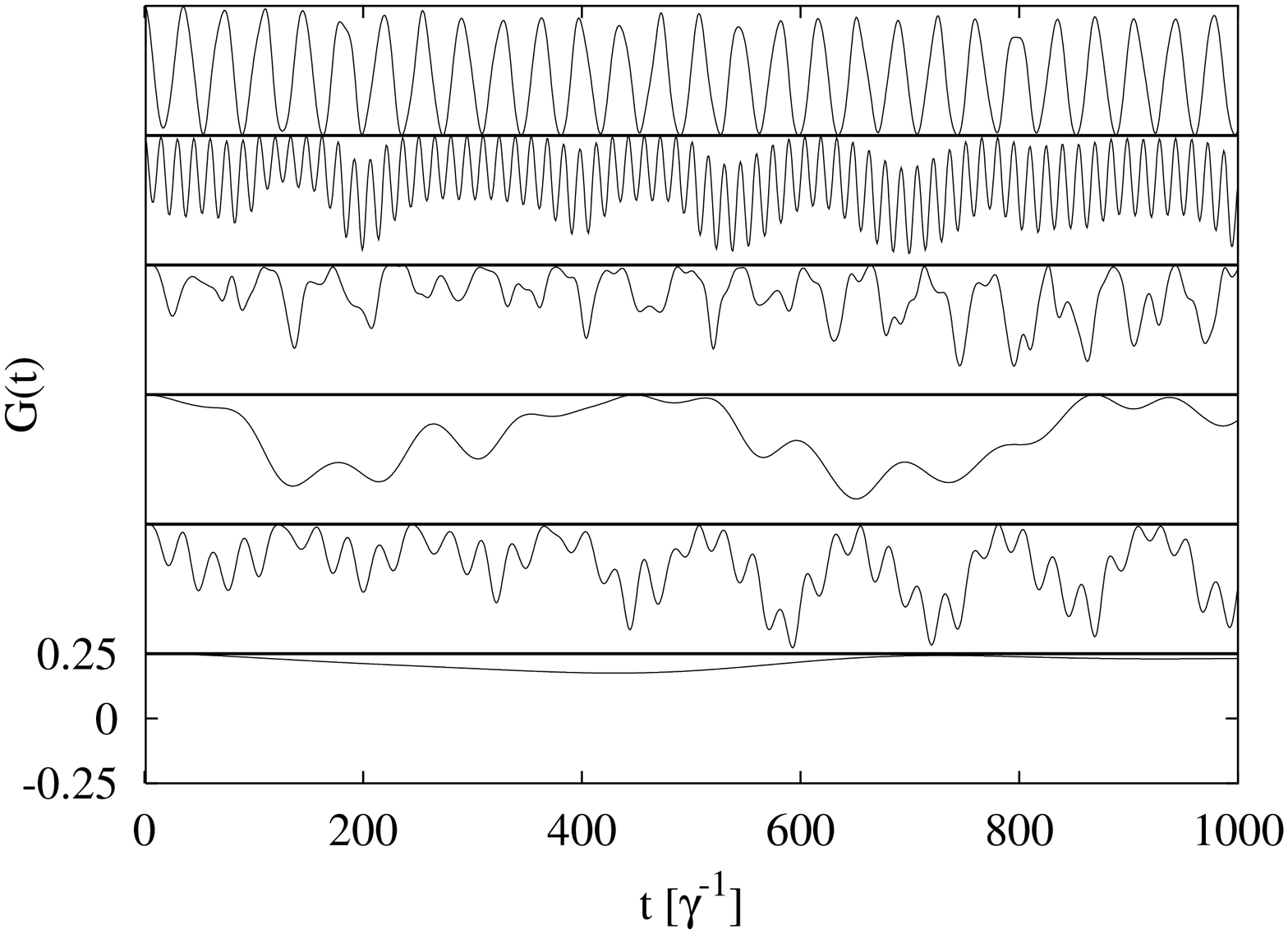}
\caption{Numerical calculations of the correlation function $G(t)$
for $N_b=32$ and various randomly chose initial conditions.  The curves are
offset for clarity and the vertical range is the same for all curves,
i.e.\ $-0.25$ 
to $0.25$} 
\label{fig:Nb32} 
\end{center}
\end{figure}

\begin{figure}[ht]
\begin{center}
\includegraphics[angle=-90,width=8cm]{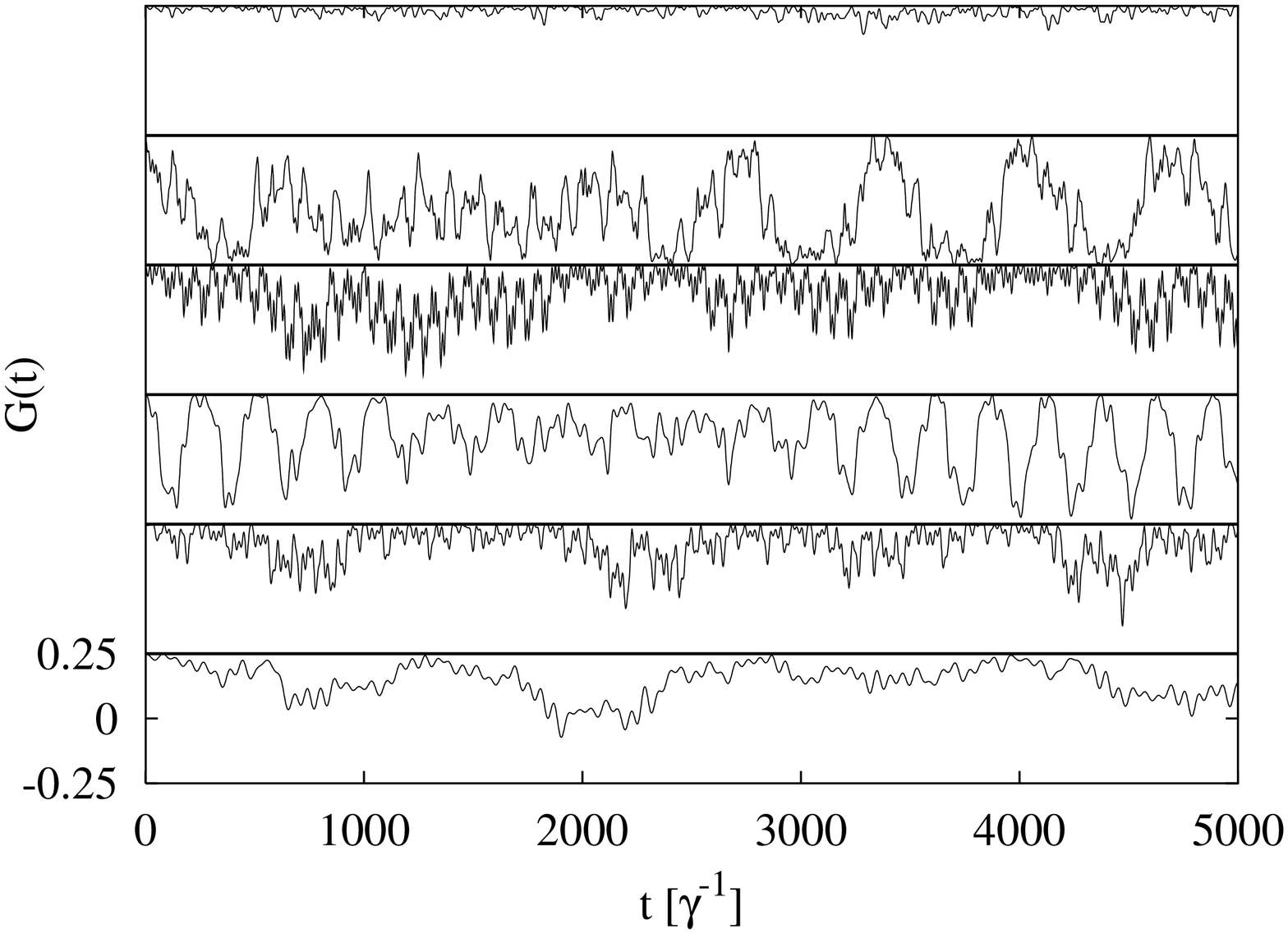}
\caption{Numerical calculations of the correlation function $G(t)$
for $N_b=128$ and various randomly chose initial conditions.  The curves are
offset for clarity and the vertical range is the same for all curves,
i.e.\ $-0.25$ 
to $0.25$} 
\label{fig:Nb128} 
\end{center}
\end{figure}

\begin{figure}[hb]
\begin{center}
\includegraphics[angle=-90,width=8cm]{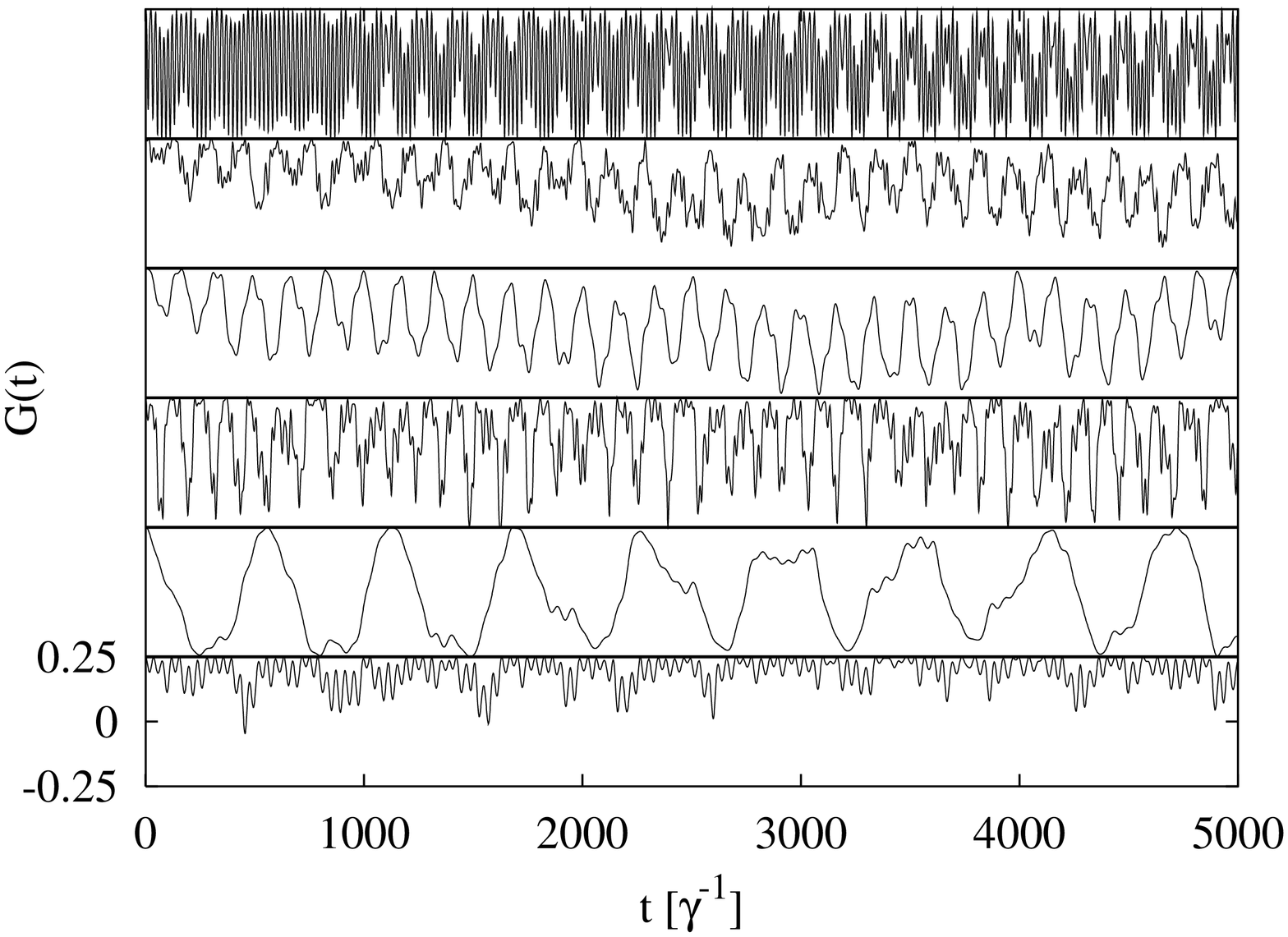}
\caption{Numerical calculations of the correlation function $G(t)$
for $N_b=256$ and various randomly chose initial conditions.  The curves are
offset for clarity and the vertical range is the same for all curves,
i.e.\ $-0.25$ 
to $0.25$} 
\label{fig:Nb256} 
\end{center}
\end{figure}

The correlation functions for a single system, i.e.\ without taking
the ensemble average, yield quite different results.  
The calculations are performed for $N_b=8,32,128$ and
$256$. For each number of subsystems $N_b$ the calculations
were repeated for various random initial conditions, but no
averaging is performed.  The results of the
calculations for $G(t)$ are presented in Figs.\
\ref{fig:Nb8}-\ref{fig:Nb256} (note different range on the horizontal,
or $t$-axis). 
The common feature of all the curves, for all values of $N_b$, is that
they do not decay with time.  This behavior persists to even
longer times, not shown here.  
Even though more complicated behavior is observed for large $N_b$, 
the characteristic frequencies of the correlation function 
oscillations do not show any obvious dependence on the number of
subsystems.

\begin{figure}[hb]
\begin{center}
\includegraphics[angle=-90,width=8cm]{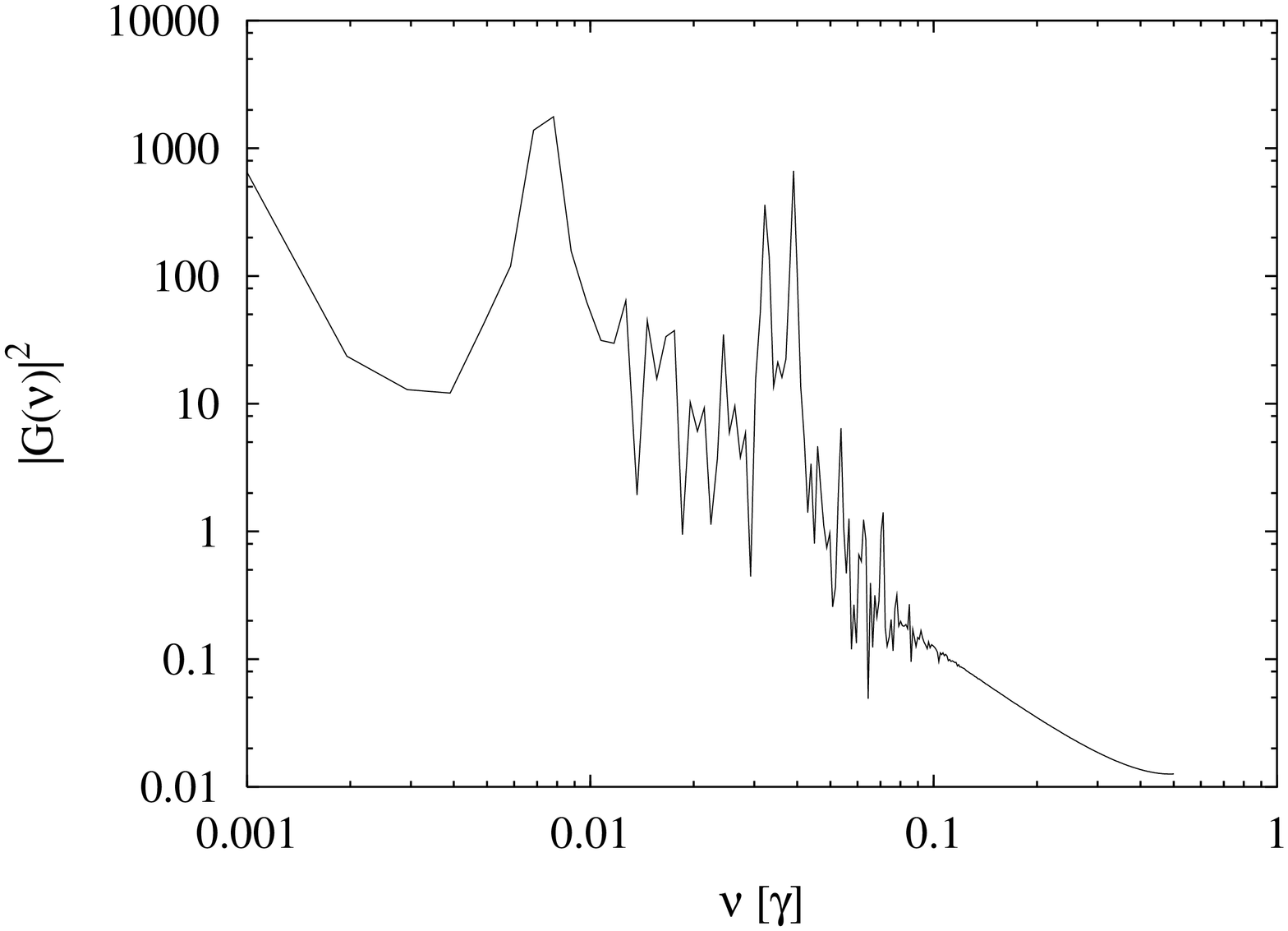}
\caption{The power spectrum of the 2nd curve (counted from the bottom 
one) in Fig.\ \ref{fig:Nb32}, corresponding to $N_b=32$.}
\label{fig:powNb32_2} 
\end{center}
\end{figure}

\begin{figure}[hb]
\begin{center}
\includegraphics[angle=-90,width=8cm]{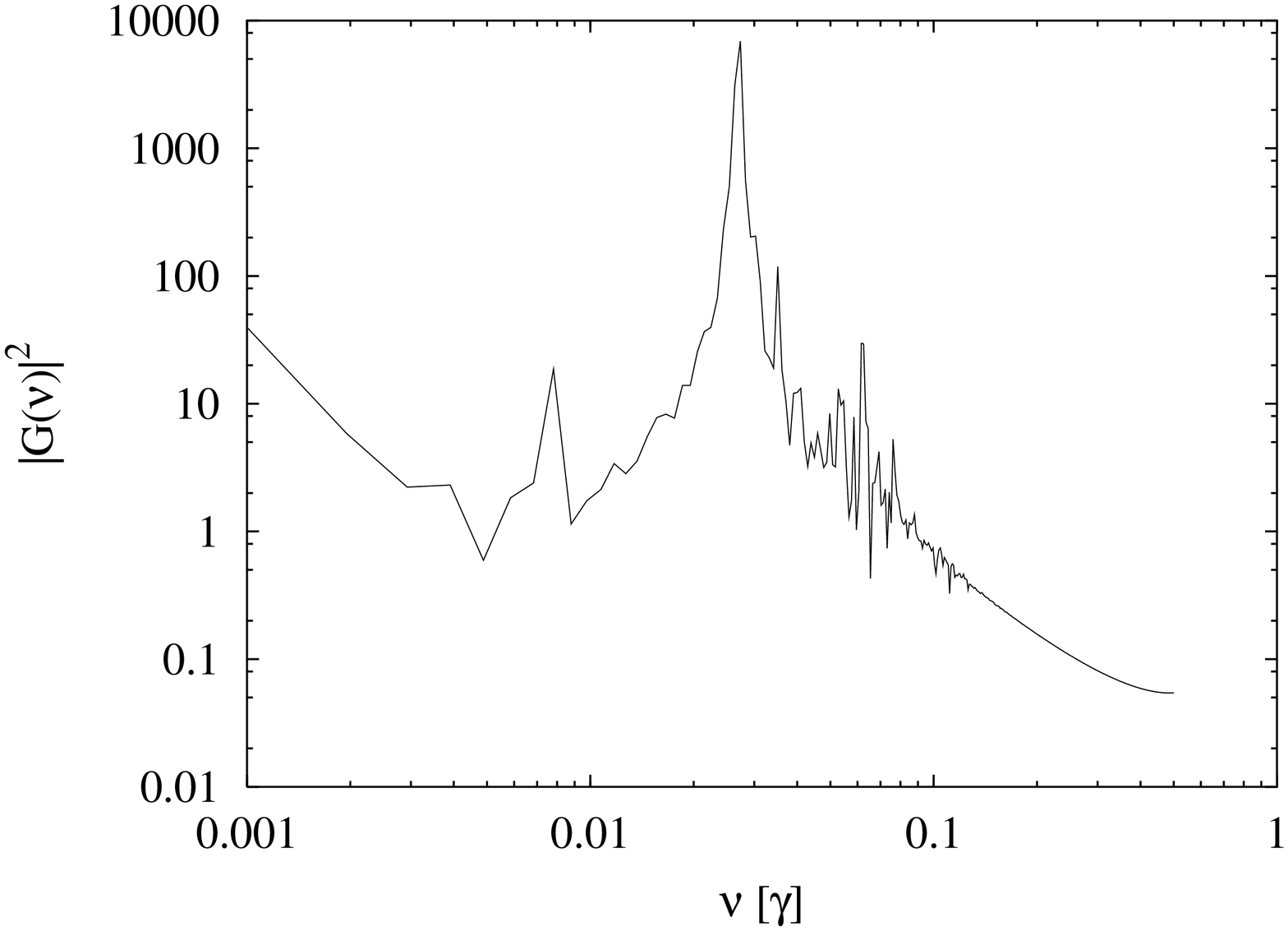}
\caption{The power spectrum of the 6th curve (counted from the bottom 
one) in Fig.\ \ref{fig:Nb32}, corresponding to $N_b=32$.}
\label{fig:powNb32_6} 
\end{center}
\end{figure}

\begin{figure}[hb]
\begin{center}
\includegraphics[angle=-90,width=8cm]{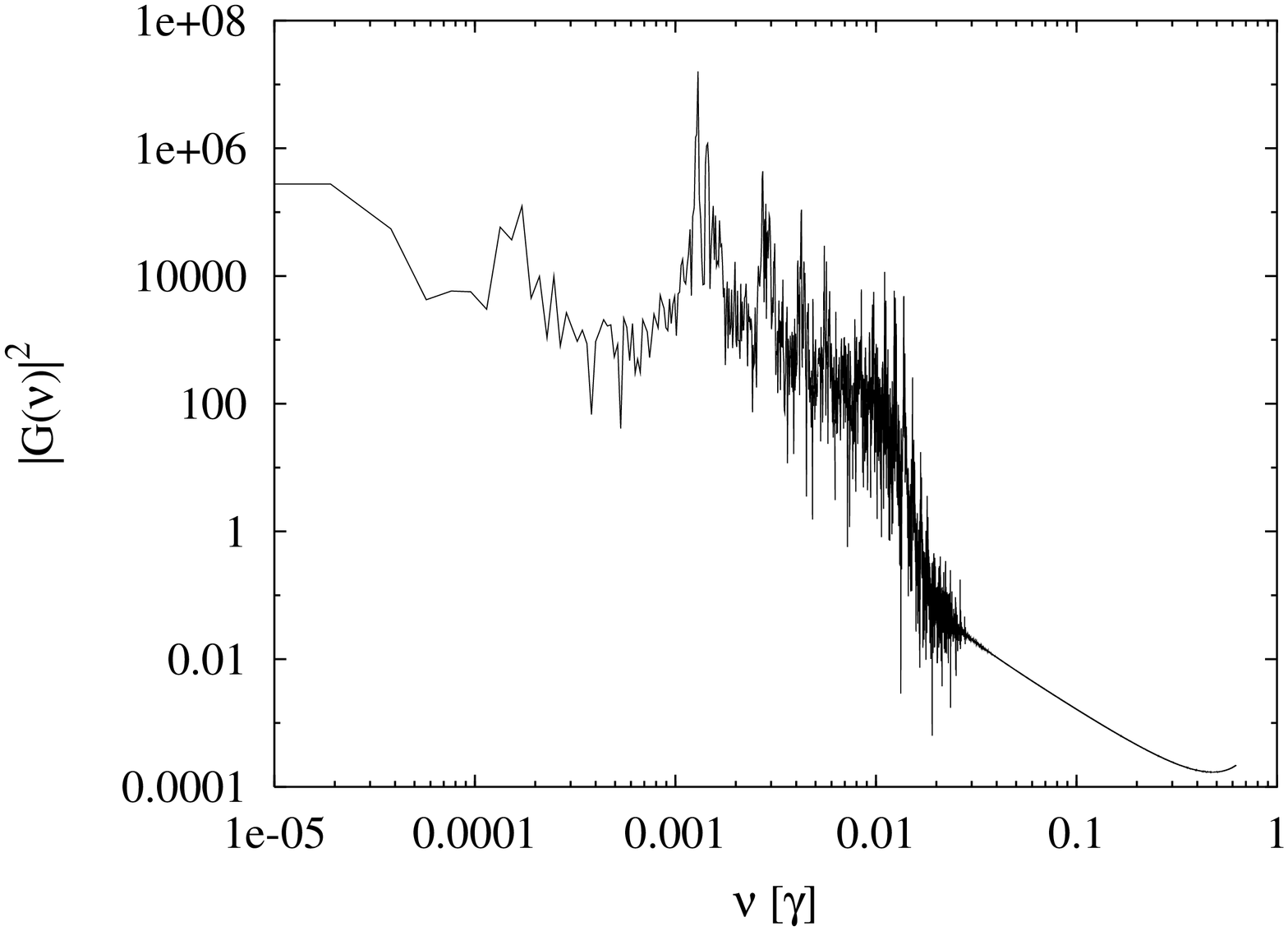}
\caption{The power spectrum of the 2nd curve (counted from the bottom 
one) in Fig.\ \ref{fig:Nb256}, corresponding to $N_b=256$.}
\label{fig:powNb256_2} 
\end{center}
\end{figure}

\begin{figure}[hb]
\begin{center}
\includegraphics[angle=-90,width=8cm]{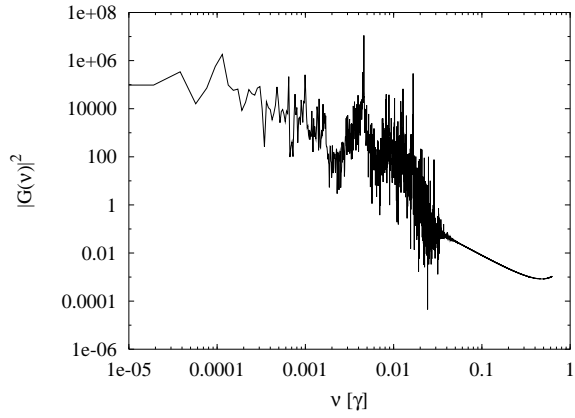}
\caption{The power spectrum of the fourth lowest curve in Fig.\
\ref{fig:Nb256}, corresponding to $N_b=256$}
\label{fig:powNb256_4} 
\end{center}
\end{figure}

\begin{figure}[hb]
\begin{center}
\includegraphics[angle=-90,width=8cm]{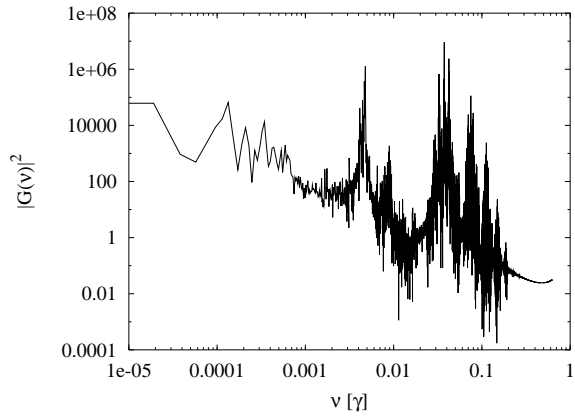}
\caption{The power spectrum of the top curve in Fig.\
\ref{fig:Nb256}, corresponding to $N_b=256$}
\label{fig:powNb256_6} 
\end{center}
\end{figure}

It is instructive to look at the power spectrum
of $G(t)$, where the oscillating
behavior of the correlation functions becomes more apparent.  
The power spectrum
(or the squared Fast Fourier Transform) of $G(t)$ is shown for the
1st and 6th curves (counted from the bottom one) for $N_b=32$ (see
Fig.\ \ref{fig:Nb32}), in Figs.\ 
\ref{fig:powNb32_2} and  \ref{fig:powNb32_6} respectively.  
The sharp, isolated peaks in the spectra illustrate well
the multi-periodic oscillations observed in $G(t)$.
This behavior is still present in the power spectra for $N_b=256$.
Figs.\ \ref{fig:powNb256_2}-\ref{fig:powNb256_6},
corresponding to respectively the 2nd, 4th and 6th curves in Fig.\
\ref{fig:Nb256}, show that even for such a complicated system (256
coupled, non-linear differential equations) the correlation functions
still show sharp, isolated peaks corresponding to well defined
oscillation periods and additional many smaller, closely spaced peaks.

The power spectrum is the Fourier spectrum of the {\em time
averaged} correlation functions, which are completely different from
the power spectra expected for the ensemble averaged correlation
functions shown in Fig.\ (\ref{fig:aveCorr}).  This implies that the time
average and the ensemble average are not equivalent, i.e.\ the system
in question is not ergodic.  The simplest way to think about this is to
consider the integrals of motion for the system.
In the case of the time-averaging the motion of the system is
at all times `constrained' by the integrals of motion, resulting in
multi-periodic correlation functions that show no decay in time
\cite{schliemann02:prc}.  
For the ensemble case, the averaged correlation 
function get contributions from many `systems' which have different
values of the integrals of motion that results in an effective
cancellation of periodic oscillations, leading to a decay of the
correlation functions.

The behavior of the correlation functions for the single system may be
broadly explained in this way: 
The oscillations of the correlation function reflect that
the system is in some sense close to being exactly solvable.   
These features will probably vanish if further terms are included into the
Hamiltonian in Eq.\ (\ref{eq:hamiltonian}).  The most
natural term would be the dipole-dipole interaction between the nuclei
which would kill most of the integrals of motion.  It is 
important to recognize  that the timescale related to the
dipole-dipole interaction is very long, of the order $10^{-3}$\,s, and
the resulting decay time would reflect that.

In connection to coherently controlling the spin,
the motion of the electron spin in the effective nuclear magnetic
field will cause `errors'.  Even though in this model nuclear spins do
not decohere the electron spin (in the sense of not causing decay of
correlation functions as a function of time), they leads to a
complicated, and unpredictable,  evolution.
Consequently, an electron spin initially in $|\uparrow
\rangle$ can be found in the opposite spin state on a timescale
$\propto \gamma^{-1}$.  Thus, even though the hyperfine coupling to
the nuclear system does not lead to decoherence in this model it
can strongly affect the dynamics of the electron spin.  

The authors acknowledge financial support from FOM and SIE would like to
thank Oleg Jouravlev, Dmitri Bagrets and Lieven Vandersypen for
fruitful discussions.  
\appendix
\section{The variance in terms of $g_b$}
\label{app:variance}
The initial condition for each nuclear spin subsystem is chosen from a
Gaussian distribution whose variance is determined by 
\begin{eqnarray}
\langle \bm{K}_b^2 \rangle 
&=&\gamma^2 I(I+1) g_b^2 C_n V_b.
\end{eqnarray}
where $V_b$ is the volume of subsystem $b$, and $C_n V_b$ is the
associated number of nuclear spins in that volume. 
For the calculations it
is convenient to express the variance for a given subsystem only in
terms of the coupling $g_b$ by expressing the subsystem volume $V_b$ as
a function of $g_b$.  
The volume of the subsystem $V_b$ is related to $g_b$ via 
\begin{eqnarray}
V_b &\equiv& \left | \frac{dV}{dg}\right |_{g=g_b} \delta g ,
\end{eqnarray}
where $V$ is the volume of the region where $g \ge g_b$.
The functional form of $g(\bm{r})$ is determined by the density
$|\psi(\bm{r})|^2$.   
In the numerical calculations
a lateral parabolic confinement of an underlying 2DEG is used.
Assuming  a constant electron density in the $z$ direction (growth
direction) the electron density is
\begin{equation}
g(\bm{r})=\theta(z_0/2-|z|)\exp(-(x^2+y^2)/\ell^2),
\end{equation}
which gives the simple relation $x^2+y^2=\ell^2 \ln(1/g)$, within the
2DEG.  Using the relation for the volume $V(\bm{r})=z_0 \pi (x^2+y^2)$ 
the subsystem volume is
\begin{eqnarray}
V_b&=& V_\mscr{QD}\frac{\delta g}{g_b},
\end{eqnarray}
which gives the variance of the distribution of the effective
nuclear magnetic field for a given subsystem in terms of the coupling 
\begin{eqnarray}
\langle \bm{K}_b^2 \rangle 
&=&\gamma^2 N I(I+1)g_b \delta g . 
\end{eqnarray}

These results can be used to calculate the variance of the total
nuclear spin defined in Eq.\ (\ref{eq:I})
\begin{eqnarray}
\langle \bm{I}^2 \rangle &=&\sum_{b,b'}
\frac{\langle \bm{K}_{b} \cdot \bm{K}_{b'} \rangle}{g_b g_{b'}} \\
&=&\frac{\langle \bm{K} \rangle}{\overline{g^2}} \sum_b \frac{\delta
g}{g_b} \\
&\approx&
\frac{\langle \bm{K} \rangle}{\overline{g^2}} \ln(2N_b),\quad N_b
\gg 1.  
\label{eq:rootMeanSquareI}
\end{eqnarray}
In the last step it is assumed that $g_b=1-(b-1/2)/N_b$, resulting in
the logarithm.

%
%

%
%



\end{document}